# *Machine Learning based Parameter Sensitivity of Regional Climate Models – A Case Study of the WRF Model for Heat Extremes over Southeast Australia*

P. Jyoteeshkumar Reddy[1, *], Sandeep Chinta[2], Richard Matear[1], John Taylor[3], Harish Baki[4], Marcus Thatcher[5], Jatin Kala[6, 7], Jason Sharples[7, 8, 9]

[1] *Commonwealth Scientific and Industrial Research Organisation Environment, Hobart, TAS, Australia*
[2] *Center for Global Change Science, Massachusetts Institute of Technology, Cambridge, MA, USA*
[3] *Commonwealth Scientific and Industrial Research Organisation Data61, Canberra, ACT, Australia*
[4] *Faculty of Civil Engineering and Geosciences, TU Delft, Delft, The Netherlands*
[5] *Commonwealth Scientific and Industrial Research Organisation Environment, Aspendale, VIC, Australia*
[6] *Environmental and Conservation Sciences, Centre for Climate-Impacted Terrestrial Ecosystems, Harry Butler Institute, Murdoch University, Murdoch, WA, Australia*
[7] *Australian Research Council Centre of Excellence for Climate Extremes, University of New South Wales, NSW, Australia*
[8] *School of Science, University of New South Wales Canberra, ACT, Australia*
[9] *NSW Bushfire and Natural Hazards Research Centre, NSW, Australia*
[*] *Corresponding author: P. Jyoteeshkumar Reddy ([jyoteesh.papari@csiro.au](jyoteesh.papari@csiro.au))*

## *Abstract*

Heatwaves and bushfires cause substantial impacts on society and ecosystems across the globe. Accurate information of heat extremes is needed to support the development of actionable mitigation and adaptation strategies. Regional climate models are commonly used to better understand the dynamics of these events. These models have very large input parameter sets, and the parameters within the physics schemes substantially influence the model's performance. However, parameter sensitivity analysis (SA) of regional models for heat extremes is largely unexplored. Here, we focus on the southeast Australian region, one of the global hotspots of heat extremes. In southeast Australia Weather Research and Forecasting (WRF) model is the widely used regional model to simulate extreme weather events across the region. Hence in this study, we focus on the sensitivity of WRF model parameters to surface meteorological variables such as temperature, relative humidity, and wind speed during two extreme heat events over southeast Australia. Due to the presence of multiple parameters and their complex relationship with output variables, a machine learning (ML) surrogate-based global sensitivity analysis method is considered for the SA. The ML surrogate-based Sobol SA is used to identify the sensitivity of 24 adjustable parameters in seven different physics schemes of the WRF model. Results show that out of these 24, only three parameters, namely the scattering tuning parameter, multiplier of saturated soil water content, and profile shape exponent in the momentum diffusivity coefficient, are important for the considered meteorological variables. These SA results are consistent for the two different extreme heat events. Further, we investigated the physical significance of sensitive parameters. This study's results will help in further optimising WRF parameters to improve model simulation.

# 1 Introduction

Extreme weather events are occurring more frequently and intensely in recent years compared to the past across the globe (Masson-Delmotte et al., 2021). Among these extreme events, heatwaves and bushfires are challenging to society due to their substantial and persistent impact on human and natural systems (Abram et al., 2021; Perkins-Kirkpatrick and Lewis, 2020). Recent studies show that heatwaves are intensifying globally and regionally in Australia, underscoring the urgent need for efficient mitigation and adaptation techniques (Reddy et al., 2021; Perkins-Kirkpatrick and Lewis, 2020). Due to its frequent exposure to heatwaves, bushfires, and other extreme events, Southeast Australia, in particular, has been identified as a region with a greater susceptibility to the effects of climate change (Bureau of Meteorology and CSIRO, 2020; Di Virgilio et al., 2019a; Sharples et al., 2016). The increasing frequency and intensity of bushfires in the region have been linked to climate change, further aggravating the risks and consequences of these events (Canadell et al., 2021; Sharples et al., 2016). Considering these events have disastrous impacts, accurately simulating heat extremes using numerical weather prediction (NWP) models is critical to support appropriate planning of adaptation and mitigation efforts.

The accuracy of a simulation from an NWP model depends on the exactness of the initial conditions and the ability of the model to represent the physical processes that occur in the atmosphere (Bjerknes, 1910). These physical processes occur at different spatial and temporal scales. Due to limited computational resources, the model cannot accurately represent the physical processes at all scales. Certain physical processes occur at smaller scales than those captured by the model grid. Moreover, even at scales equivalent to the grid resolution, some processes have complex mechanisms that are computationally intensive or intricate to be represented explicitly. Parameterizations address this shortcoming by offering simplified, computationally efficient representations of complex physical processes, such as convection, cloud formation, soil moisture transport, and boundary layer physics, to name a few. Parameterization schemes employ mathematical equations to capture these complex physical processes and relate these partially represented processes to the large-scale processes that the model resolves. These equations are typically based on empirical relationships or theoretical understanding derived from observations or detailed process studies (Hong and Lim, 2006; Kain, 2004). Numerous tuneable parameters are frequently included in the parameterization schemes to simulate atmospheric processes. These parameters are typically constants and exponents in the model equations, and their default values are specified by the scheme developers based on experimental or theoretical investigations (Hong et al., 2004). Therefore, accurate specification of these parameters within the model's physical representation is crucial for the reliability of model simulations (Di et al., 2015; Yang et al., 2012).

The Weather Research and Forecasting (WRF) model is an NWP model widely used as a weather forecasting tool and for dynamical downscaling purposes across the globe (Evans et al., 2014; Skamarock, 2008). Researchers and meteorological organizations have adopted the model extensively due to its adaptability, flexibility, and high-resolution capabilities (Evans et al., 2014; Skamarock, 2008). The WRF model has been used in numerous studies in simulating extreme events like heatwaves and extreme bushfires in Southeast Australia (Di Virgilio et al., 2019a; Kala et al., 2023, 2015). The WRF model has multiple options for parameterization schemes for each physical process. Different combinations of parameterization schemes give the user a choice of model configurations (Skamarock, 2008). Previous studies have examined the sensitivity of various physics parameterization schemes in simulating extreme events over Southeast Australia (Evans et al., 2014, 2012; Ji et al., 2022) but no studies have explored the sensitivity of parameters within the physics schemes. As discussed earlier, the values of parameters in these physics parameterization schemes play a crucial role in model fidelity. Therefore, calibration of these parameter values using observations could help improve the model simulation (Baki et al., 2022a; Chinta and Balaji, 2020; Duan et al., 2017).

Parameter estimation can be a complex process due to several challenges. With many tuneable parameters (often in tens to hundreds), the number of model simulations required to calibrate these parameters increase exponentially as the number of tuneable parameters increases. Furthermore, multiple meteorological variables, such as humidity, wind speed, and temperature, to name a few, must be considered simultaneously during calibration. The parameter calibration process requires enormous computational resources. This challenge can be overcome by first implementing sensitivity analysis (SA) to identify the parameters significantly influencing the model output variables of interest. This approach helps reduce the number of parameters that need calibration, ultimately improving the accuracy of the variables of interest in the model output (Chinta et al., 2021; Quan et al., 2016). Sensitivity analysis investigates how variations in the model output can be attributed to different sources of uncertainty in the model's input parameters (Saltelli, 2002). Several studies have implemented sensitivity analysis to WRF model parameters to identify the parameters that influence the output variables (Baki et al., 2022b; Di et al., 2018, 2017, 2015; Ji et al., 2018; Wang et al., 2020; Yang et al., 2012).

This study aims to identify the WRF model parameters that most affect the different model output meteorological variables related to heat extremes (i.e., heatwaves and fire weather). We do this by focussing on Southeast Australia during two extreme fire weather and heatwave events. This study is organized as follows; Section 2 introduces the surrogate model based global SA technique and describes the data, events selected, and WRF model parameters for the SA. Section 3 presents the SA results and a detailed discussion of the physical significance of identified sensitive parameters. Section 4 summarises the conclusions from this study.

## 2 Methods
### 2.1 *Machine learning surrogate-based Sobol sensitivity analysis*

Sensitivity analysis (SA) is broadly classified into local and global categories. The local SA is performed by exploring how the perturbed input parameters around specific reference values influence the ouput(s). However, global SA explores the entire feasible parameter space and highlights each parameter's total effects, including their interactive effects on the model output. The local SA method is widely used because of its limited computational requirements (Rakovec et al., 2014). However, if the input parameters interact and have a non-linear influence on the output(s), the local SA method will be substantially biased and underestimate the importance of parameters (Saltelli et al., 2019). The considered WRF model parameters are expected to have complex interactions and non-linearly influence the outputs (Baki et al., 2022b; Chinta et al., 2021; Di et al., 2015; Quan et al., 2016). Hence, in this study, we use the global SA method. There are many global SA techniques, including Morris One-at-A-Time, Multivariate adaptive regression splines, Fourier amplitude sensitivity test, and Sobol sensitivity analysis. Out of these global SA methods, previous WRF parameter SA studies recommend the Sobol SA method because of its accuracy, capacity to address interaction effects, discontinuities, and possible non-linear effects of the parameters on the output variables (Baki et al., 2022b; Wang et al., 2020).

Sobol sensitivity analysis is based on Analysis of Variance decomposition, also known as Hoeffding-Sobol decomposition, which states that a functional output variance can be expressed as the combination of variances of the output function contributed by individual parameters, interactions of parameter pairs, and so on in the increasing dimensionality (Saltelli et al., 2010; Sobol′, 2001). Based on the definition of variance as a measure of likeliness, the Sobol sensitivity indices are estimated as the ratios of variances contributed by the parameter interactions with respect to the total variance. The computation of sensitivity indices requires functional evaluations over a large number of input samples, which is computationally very expensive for the WRF model. However, it is feasible to generate the thousands of output samples required for the sensitivity analysis by employing a surrogate Machine Learning (ML) model that has been well trained on a limited set of WRF model runs. Among the various ML models available for use as surrogate models, Gaussian Process Regression (GPR) has been identified as one of the best suitable choices in such scenarios (Baki et al., 2022b; Ji et al., 2018).

GPR is a probabilistic and non-parametric machine learning technique that is accurate enough to predict and model unknown functions. The GPR is built upon the Gaussian Process, which is "a collection of random variables, any finite number of which have consistent Gaussian distributions" (Williams and Rasmussen, 1995, 2006). For any vector of random input variables $x = \{x_1, x_2, \ldots, x_n\}$, the corresponding latent functions $f(x) = \{f(x_1), f(x_2), \ldots, f(x_n)\}$ form a joint multivariate Gaussian distribution governed by a mean function and kernel function. Based on this, the GPR can be explained in four stages. First, assuming a mean and a kernel function, a collection of random output latent functions is obtained over random input variables, known as a Gaussian prior distribution. Second, provided a set of observed training samples and assuming the targets follow a Gaussian distribution, the hyperparameters of the kernel function are obtained by maximizing the log-likelihood of the target's Gaussian distribution. Third, with the optimized hyperparameters, a posterior distribution can be modelled by conditioning the observations for a new set of test points. Finally, the posterior distribution can be used to make predictions of any set of new test points and keep the model updated.

## 2.2 *WRF model configuration*

This study employs the Advanced Research WRF (WRF-ARW) model version 4.4 for the numerical simulations (Skamarock, 2008). The model domain encompassing southeast Australia consists of 206×181 grid points with a horizontal resolution of 12 km (Fig. S1). The model has 40 σ vertical levels with the top layer set at the 50 hPa atmospheric level. The time integration is performed with a time step of 72 s. The European Centre for Medium-Range Weather Forecast (ECMWF) Reanalysis 5$^{th}$ generation data set (ERA5) (Hersbach et al., 2020) at 0.25° horizontal resolution with a six-hourly interval are used as the initial and lateral boundary conditions. The WRF model output data is stored at hourly intervals. The parameters and physics schemes are adapted based on previous studies (Baki et al., 2022b; Chinta et al., 2021; Di et al., 2015; Quan et al., 2016). The similar WRF physics parametrisation scheme combination performed reasonably well in simulating the climatology and extreme temperatures over the study domain (Evans and McCabe, 2010; Kala et al., 2015). A total of 24 adjustable model parameters are identified across seven different physics schemes. These 24 parameters are chosen because they have documented ranges. The details are presented in Table 1.

*Table 1. The WRF model physics schemes used in this study with corresponding adjustable parameters in the schemes and their respective default values and ranges. These default values and ranges are adapted from previous studies* (Baki et al., 2022b; Chinta et al., 2021; Di et al., 2015; Quan et al., 2016).

| **Physics scheme** | **Parameter** | **Default** | **Range** | **Description** |
|---|---|---|---|---|
| Surface layer -MM5 Monin-Obukhov Scheme (Jiménez et al., 2012) | P1 | $2.4 \times 10^{-5}$ | $1.2 \times 10^{-5} - 5 \times 10^{-5}$ | The parameter for heat/moisture exchange coefficient (s/m$^2$) |
| | P2 | 0.0185 | $0.01 - 0.037$ | The coefficient for converting wind speed to roughness length over water |
| | P3 | 1 | $0.5 - 2$ | Scaling related to surface roughness |
| | P4 | 0.4 | $0.35 - 0.42$ | Von Kármán constant |
| Cumulus - Kain-Fritsch Eta scheme (Kain, 2004) | P5 | 1 | $0.5 - 2$ | The multiplier for downdraft mass flux rate |
| | P6 | 1 | $0.5 - 2$ | The multiplier for entrainment mass flux rate |
| | P7 | 150 | $50 - 350$ | Starting height of downdraft above USL (hPa) |
| | P8 | 2700 | $1800 - 3600$ | Average consumption time of CAPE (s) |
| | P9 | 5 | $3 - 12$ | The maximum turbulent kinetic energy value in sub-cloud layer (m$^2$/s$^2$) |
| Microphysics - WSM 6-class scheme (Hong and Lim, 2006) | P10 | 14900 | $8000 - 30000$ | Scaling factor applied to ice fall velocity (1/s) |
| | P11 | $8 \times 10^6$ | $5 \times 10^6 - 12 \times 10^6$ | Intercept parameter of rain (1/m$^4$) |
| | P12 | $5 \times 10^{-4}$ | $3 \times 10^{-4} - 8 \times 10^{-4}$ | The limited maximum value for the cloud-ice diameter (m) |
| | P13 | 0.55 | $0.35 - 0.85$ | Collection efficiency from cloud to rain auto conversion |
| Shortwave radiation - Dudhia scheme (Dudhia, 1989) | P14 | $1 \times 10^{-5}$ | $0.5 \times 10^{-5} - 2 \times 10^{-5}$ | Scattering tuning parameter (m$^2$/kg) |
| Longwave radiation - RRTM scheme (Mlawer et al., 1997) | P15 | 1.66 | $1.55 - 1.75$ | Diffusivity angle for cloud optical depth computation |
| Land surface - unified Noah land surface model scheme (Chen and Dudhia, 2001; Tewari et al., 2004) | P16 | 1 | $0.5 - 2$ | The multiplier for hydraulic conductivity at saturation |
| | P17 | 1 | $0.5 - 2$ | The multiplier for the saturated soil water content |
| | P18 | 1 | $0.5 - 2$ | The multiplier for minimum soil suction |
| | P19 | 1 | $0.5 - 2$ | The multiplier for Clapp and hornbereger "b" parameter |
| Planetary boundary layer - Yonsei University (YSU) scheme (Hong et al., 2006) | P20 | 0.3 | $0.15 - 0.6$ | Critical Richardson number for boundary layer of water |
| | P21 | 0.25 | $0.125 - 0.5$ | Critical Richardson number for boundary layer of land |
| | P22 | 2 | $1 - 3$ | Profile shape exponent for calculating the momentum diffusivity coefficient |
| | P23 | 6.8 | $3.4 - 13.6$ | Coefficient for Prandtl number at the top of the surface layer |
| | P24 | 15.9 | $12 - 20$ | Counter gradient proportional coefficient of non-local flux of momentum |

## 2.3 Experimental setup

In this study, we select the two heat extremes where heatwaves and extreme fire weather days are observed over southeast Australia for the numerical simulations. One is the southeast Australian heatwave event starting in the last week of January and extending to the second week of February 2009 (Engel et al., 2013). The 2009 event is simulated for 13 days covering the period of extreme heat, i.e., from 26$^{th}$ Jan 12 UTC to 08$^{th}$ Feb 12 UTC. Another event is the heat extremes during the mid to end of December 2019/20 period. The 2019 event is simulated for 15 days covering heatwave and extreme fire weather days, i.e., from 16$^{th}$ Dec 12 UTC to 31$^{st}$ Dec 12 UTC. A 12 h model spin-up is considered for both events. The simulation results are evaluated against the Bureau of Meteorology Atmospheric high-resolution Regional Reanalysis for Australia (BARRA2) data (Su et al., 2022). BARRA2 provides hourly data at 12 km horizontal resolution over the Australian region.

This study conducts parameter SA for the critical meteorological variables of heat extremes such as temperature, relative humidity, and wind speed during both events (2009 and 2019). Firstly, a Quasi-Monte Carlo (QMC) Sobol sequence design is employed to generate 256 parameter samples using the Uncertainty Quantification Python Laboratory (UQ-PyL) package as recommended by the previous studies (Baki et al., 2022b; Wang et al., 2020). Next, we perform the WRF simulations with the 256 generated parameter samples for each event. Then we calculate each meteorological variable's root mean square error (RMSE) between WRF simulations and BARRA2 data. The RMSE of 256 WRF runs for both the events is presented in Figure 1. The RMSE is normalised with min-max normalisation for each selected event, respectively. The 256 WRF runs with perturbed parameters are broadly consistent for both the events with minor differences which are expected due to slightly different event characteristics (Fig. 1).

We then train the GPR model with parameter samples as input and RMSE as output for each of the considered meteorological variables. The GPR model is evaluated against the WRF RMSE using a K-fold cross-validation technique. Here the entire dataset is divided into K-folds (K=8). The GPR model is trained with (K-1) folds; the left-out $K^{th}$ fold is used as test data. It is iterated over all the folds, and each test set's GPR predictions are stacked. The accuracy of the GPR model is measured with the goodness of fit ($R^2$) metric by comparing the GPR predictions with WRF data. This shows that GPR has good accuracy with high $R^2$ values for each considered meteorological variable (Fig. S2). Consequently, the trained GPR model is used for predicting the RMSE of the 50,000 parameter samples. These parameter samples are generated using the Sobol sequence design, following previous studies (Baki et al., 2022b; Wang et al., 2020). Finally, Sobol sensitivity indices are calculated based on these 50,000 samples, thereby providing insights into the relative importance of the parameters.

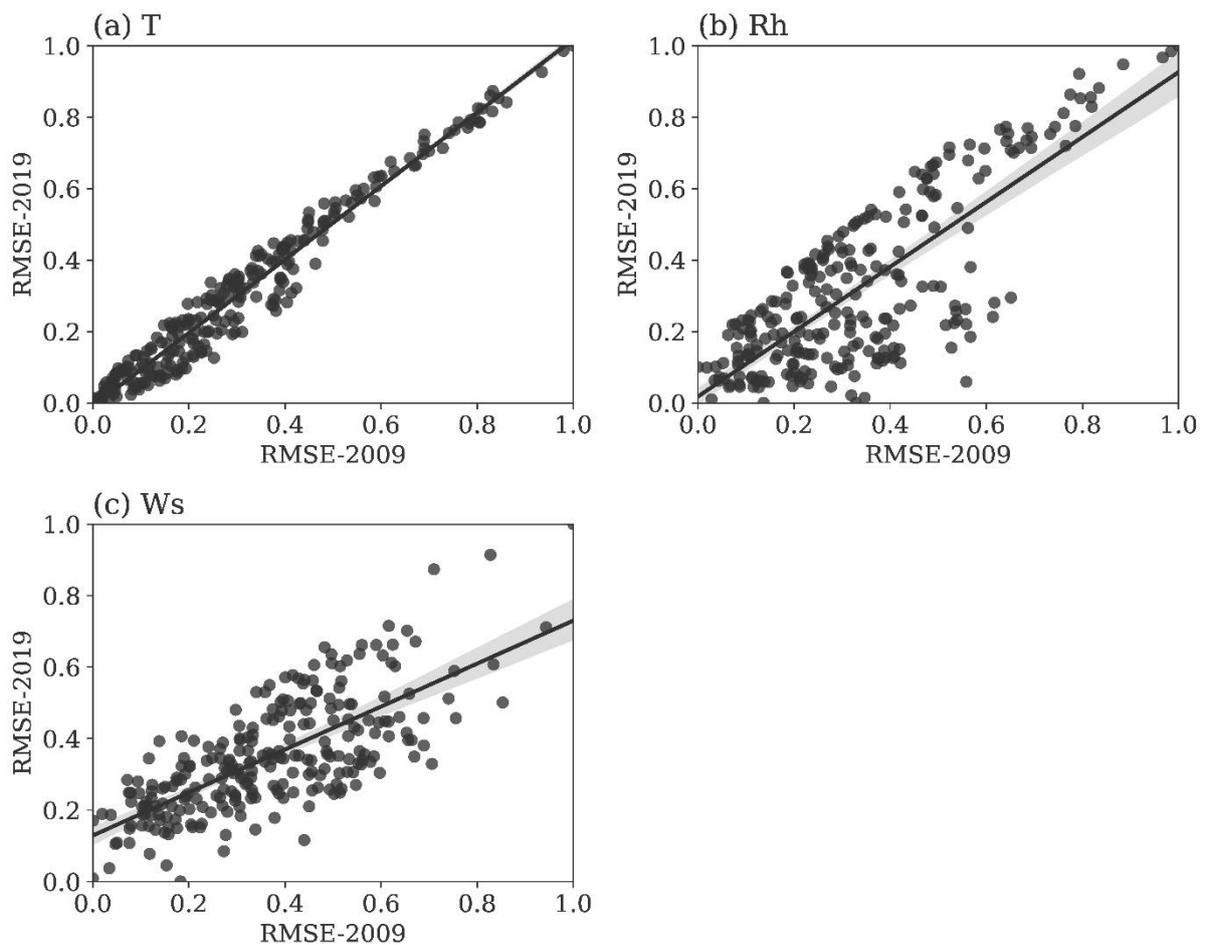

*Fig. 1 Comparison of normalised RMSE of air temperature (at 2m height) (T; °C), relative humidity (at 2m height) (Rh; %), and wind speed (at 10m height) (Ws; m/s) of 256 WRF parameter runs for both considered events (2009 and 2019). Both the events are simulated with the same set of 256 parameter values (in the same order). The line refers to the linear fit, and shading represents the 95% confidence level of the linear fit.*

# 3 Results and discussion

## 3.1 Parameter sensitivity analysis of heat extremes

The parameter sensitivity analysis (SA) is performed for the hourly surface (2m) air temperature (T), surface (2m) relative humidity (Rh), and surface (10m) wind speed (Ws) during the two case studies. Both the first- and higher-order Sobol sensitivity indices of parameters for the corresponding meteorological variables are presented in Figure 2. The first-order sensitivity indices represent the effect of variations in a single parameter, and higher-order sensitivity indices represent the interaction effects of changes in multiple parameters on the output (Saltelli, 2002; Saltelli et al., 2010; Sobol′, 2001). The SA results are presented individually for the considered extreme heat events (2009 and 2019) and for the combined data in Figure 2. It is observed that the SA results are mostly consistent across both events and also for the combined data. However, minor variations in the degree of sensitivity of the important parameters between the events were observed (Fig. 2). These variations in the magnitude of sensitivity could be due to the slightly different event characteristics (Fig. 1).

SA results suggest that parameters P14 (scattering tuning parameter), P17 (multiplier of saturated soil water content), and P22 (profile shape exponent in the momentum diffusivity coefficient) most affect the considered meteorological variables (T, Rh, and Ws). For T, only two parameters, P14 and P17, are most important, and their sensitivity is mostly first-order. Their higher-order sensitivity is minor compared to the first-order effects on T (Fig. 2(a)). This means that P14 and P17 individually affect the T to a large extent, and the higher-order effects are mostly due to the interactions between them (not shown). Similar to T, P14 and P17 are the most influential parameters for Rh, and in addition to these, P22 has some minor influence on Rh. Again, for Rh, the sensitive parameters (P14, P17, and P22) influence is mostly first-order (Fig. 2(b)). In contrast to T and Rh, for Ws, P22 is equally or slightly more important as the P14 and P17 parameters (Fig. 2(c)). Results show that the higher-order sensitivity of P14 and P17 for Ws is considered to be as important as first-order sensitivity. This higher-order sensitivity of P14 and P17 for Ws is partly due to their second-order interactions (not shown). Several other parameters, P6, P10, P12, P21, P23, and P24, show very minor effects on Ws (Fig. 2(c)).

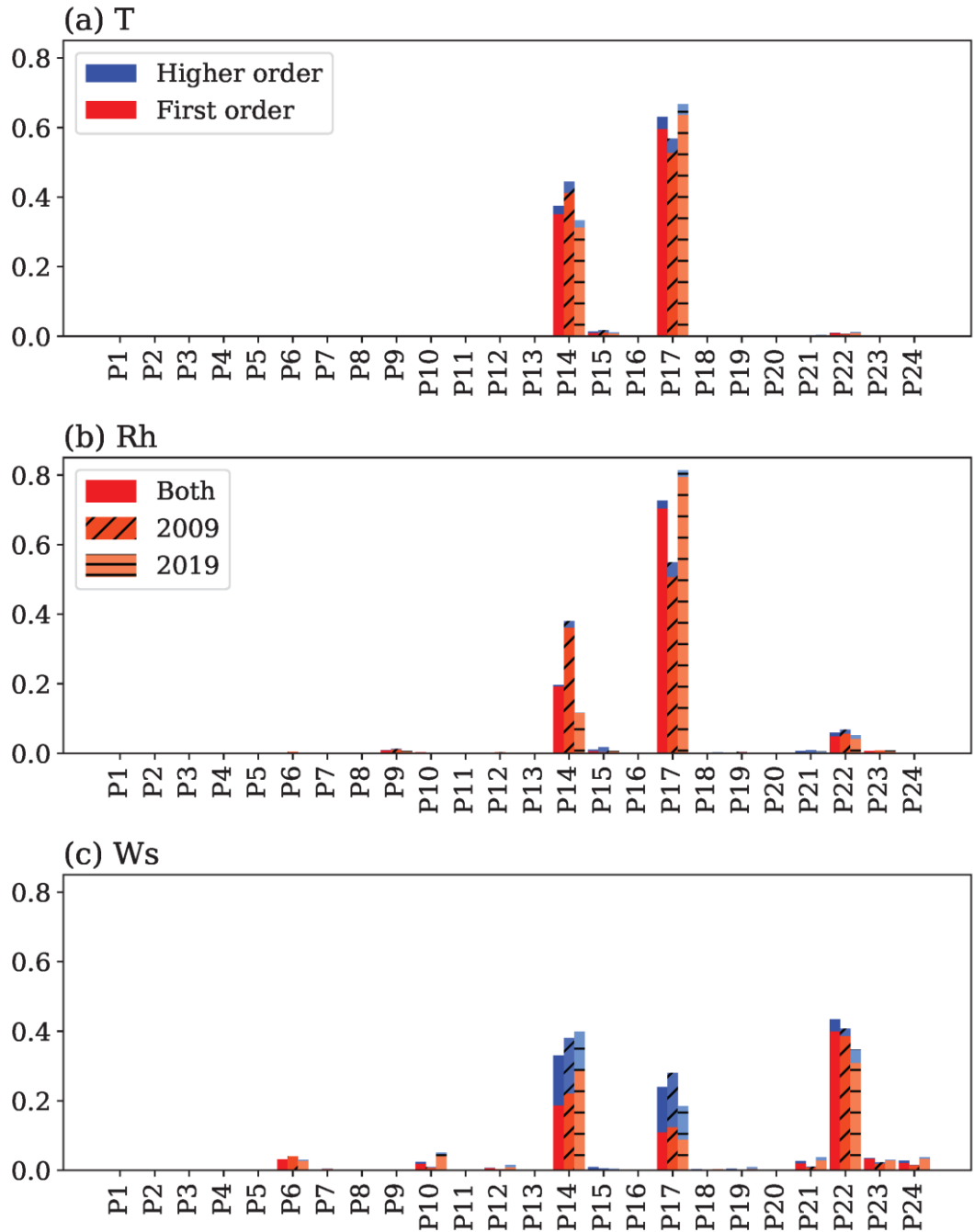

*Fig. 2 First-order (red) and higher-order (blue) sensitivity of various considered parameters (P1-P24) for (a) air temperature (at 2m) (T; °C), (b) relative humidity (at 2m) (Rh; %), and (c) wind speed (at 10m) (Ws; m/s) during the 2009 (cross stripes), 2019 (horizontal stripes) events separately and both events combined (no stripes). Total sensitivity is the sum of the first-order (red) and higher-order (blue) sensitivities.*

### 3.2  *The physical significance of sensitive parameters*

Parameter P14 is the scattering tuning parameter in the Dudhia shortwave radiation scheme, which is related to the scattering under clear sky conditions. This parameter directly influences the incoming shortwave radiation reaching the Earth's surface by affecting the scattering attenuation (Dudhia, 1989; Montornès et al., 2015). The solar radiation is inputted to the land surface parameterisation scheme, which then transforms to surface energy fluxes (sensible and latent heat). The surface fluxes regulate the mixing in the planetary boundary layer through thermals, affecting the horizontal wind speed (Oke, 2002). Therefore, T, Rh, and Ws are sensitive to P14. When P14 is low, scattering is attenuated, leading to an increase in incoming shortwave radiation, which then heats the surface and increases the surface temperature. Our results are consistent with this: as P14 decreases, the mean T values increase (Fig. 3(a)). This increase in mean T is mainly due to the increase in daily maximum temperature ($T_{max}$) (Fig. 3(b)). P14's influence on daily minimum temperature ($T_{min}$) is very minimal because the parameter is related to the scattering of shortwave radiation, which is absent during night time (Fig. 3(c)). Similar to T, P14 influences Rh (Fig. 3(d-f)). As P14 decreases, the mean Rh and mean daily minimum relative humidity ($Rh_{min}$) decrease; however, the mean daily maximum relative humidity ($Rh_{max}$) is mostly unaffected (Fig. 3(d-f)). For Ws, lower P14 results in stronger winds (both the mean and daily maximum wind speed ($Ws_{max}$)) and vice versa (Fig. 3(g, i, j, and l)).

Parameter P17, the multiplier of saturated soil water content or soil porosity in the Unified Noah land surface scheme, is the other most influential parameter for T, Rh, and Ws. P17 regulates the transmission rate of moisture and heat transport in the soil (Chen and Dudhia, 2001; Tewari et al., 2004). This implicitly affects the rate of heat transfer and water vapour exchange between the land surface and the atmosphere (Chinta et al., 2021; Tewari et al., 2004). Smaller values of P17 implies less porosity, which means more dense soil with high soil thermal inertia. The high thermal inertia means lower day time temperatures and higher night time temperatures (Fonseca et al., 2019; Quan et al., 2016; Temimi et al., 2020). This is due to an increase in conduction which means faster heat transfer between the surface and subsurface, leading to a slower rate of surface warming (cooling) during the day (night). The slow rate of day time (night time) surface warming (cooling) results in low (high) daily maximum (minimum) temperatures. Our results are consistent with this and show that at low P17, mean $T_{max}$ is lower and mean $T_{min}$ is higher, compared to those during high P17 (Fig. 3(b-c)). Similar to $T_{max}$ and $T_{min}$, as P17 decreases, the mean $Rh_{min}$ increases; however, the mean $Rh_{max}$ decreases (Fig. 3(e-f)). Overall, low P14 and low P17 results in high mean T and low mean Rh (Fig. 3(a and d)).

The parameter P22 is the profile shape exponent associated with the calculation of the momentum diffusivity coefficient in the boundary layer scheme (Hong et al., 2006). P22 governs the turbulent mixing in the boundary layer affecting the height of the maximum diffusivity and leading to the influence of the horizontal wind speed (Baki et al., 2022b; Di et al., 2015; Oke, 2002). As P22 increases, the height of momentum diffusivity declines and results in increasing (decreasing) momentum diffusivity coefficient below (above) the maximum height. This means high P22 can lead to weak winds in low levels and strong winds in the upper levels of the atmosphere, respectively (Yang et al., 2019). Our results show that lower P22 values are associated with stronger mean wind speeds (both Ws and $Ws_{max}$) and vice versa (Fig. 3 (h, i, k, and l)).

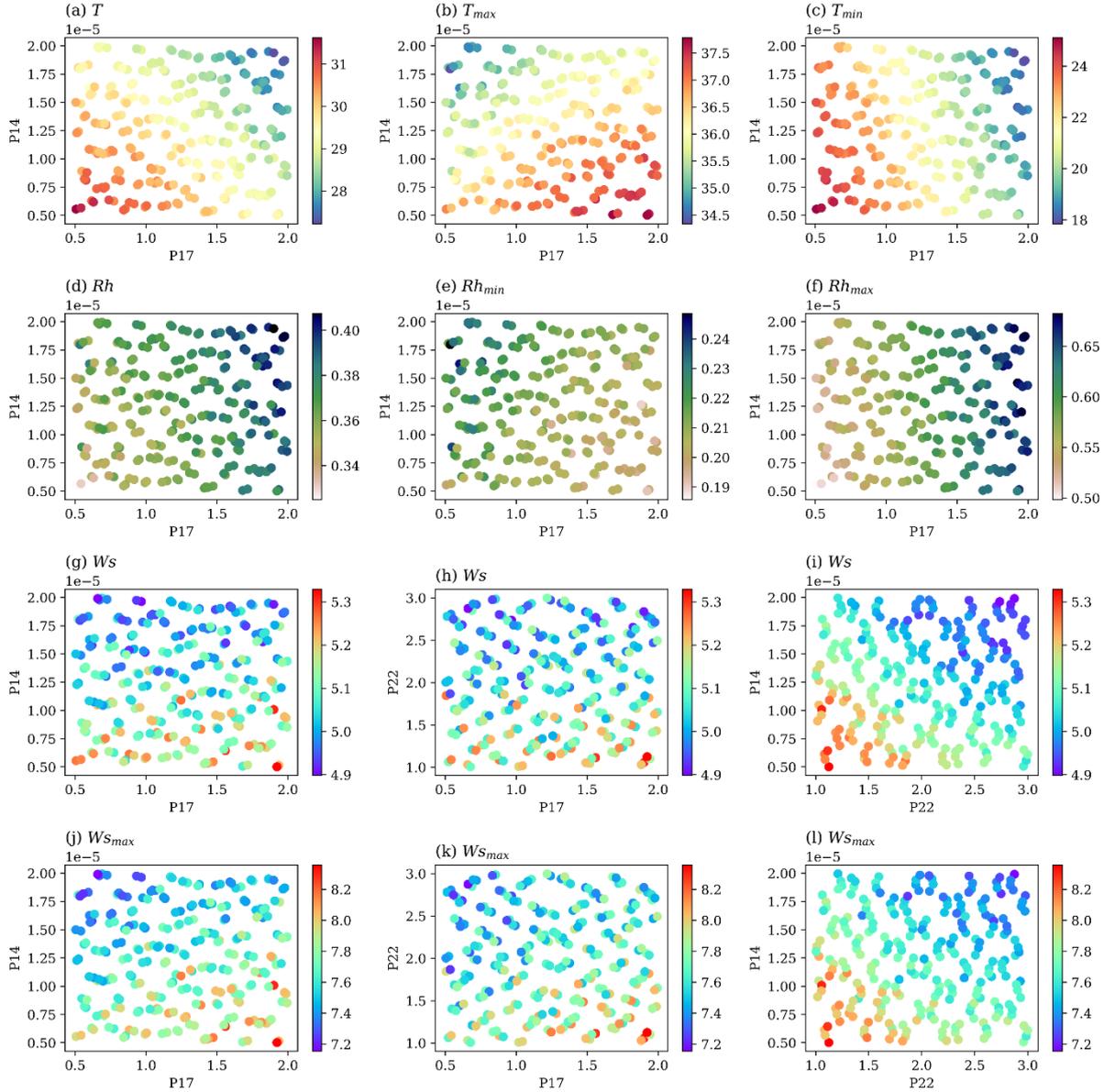

*Fig. 3 The mean of surface meteorological variable ((a) hourly temperature (T; °C), (b) daily maximum temperature ($T_{max}$; °C), (c) daily minimum temperature ($T_{min}$; °C), (d) hourly relative humidity (Rh), (e) daily minimum relative humidity ($Rh_{min}$), (f) daily maximum relative humidity ($Rh_{max}$), (g-i) hourly wind speed (Ws; m/s), and (j-l) daily maximum wind speed ($Ws_{max}$; m/s)) of 256 WRF runs with a different set of parameter values for the combined data of both events. The x- and y-axis represent the respective sensitive parameters of corresponding meteorological variables (parameters P17 and P14 for temperature (a-c), relative humidity (d-f), and wind speed (g and j); P17 and P22 for wind speed (h and k); P22 and P14 for wind speed (i and l)). The color represents the mean values of the respective meteorological variables.*

### 3.3 Comparing the default, best, and worst parameter combination results

The SA results show that only three parameters most affect the considered meteorological variables. These three sensitive parameters can be further optimised to improve WRF model simulation of critical meteorological variables for heat extremes such as $T_{max}$, $Rh_{min}$, and $Ws_{max}$.

The calibration of sensitive parameters with a multi-objective optimisation technique is a complicated task, which is outside the present study's scope but can be done in a future study. However, the 256 WRF simulations for the SA can provide us with an understanding of how different parameter sets, compared to the default values, can improve (or amplify) the model biases of the selected output variables. For this purpose, we compare the results of default, best, and worst parameter combinations out of 256 WRF runs, as illustrated in Figure 4. The parameter set with minimum (maximum) RMSE of corresponding hourly meteorological variable data (T, Rh, and Ws) out of 256 WRF runs is chosen as best (worst).

Figure 4 presents the comparison of average $T_{max}$, $Rh_{min}$, and $Ws_{max}$ during all days of both events (2009 and 2019) for the default, best, and worst parameter sets with respect to BARRA2 data. The default simulation results suggest that $T_{max}$ is under-predicted by around 2 – 4 °C and over-predicted both the $Rh_{min}$ (by about 0.06) and $Ws_{max}$ (by approximately 2 – 4 m/s) compared to the BARRA2 data over much of the study domain. These changes are more substantial over the eastern coastal strip (Fig. 4 (b, f, and j)). These results are broadly consistent with the previous studies, which showed the cold biases in the default parameter WRF model simulated $T_{max}$ compared to observations over the study region (Di Virgilio et al., 2019b; Ji et al., 2022; Kala et al., 2015). However, the best parameter combination results suggest that the simulation of $T_{max}$, $Rh_{min}$, and $Ws_{max}$ is improved over much of the study domain compared to the default run (Fig. 4 (c, g, and k)). The improvements can be clearly seen over the regions of $T_{max}$ greater than 40 °C (compare Fig. 4 (b) and (c)). Conversely, $T_{max}$, $Rh_{min}$, and $Ws_{max}$ biases are amplified for the worst parameter set compared to the default simulation (Fig. 4 (d, h, and l)). These results are consistent even when only the extremely hot day of each event is considered separately (Figs. S3 and S4). Overall, these results suggest that tuning the most sensitive parameters can further improve model simulation of meteorological variables of interest.

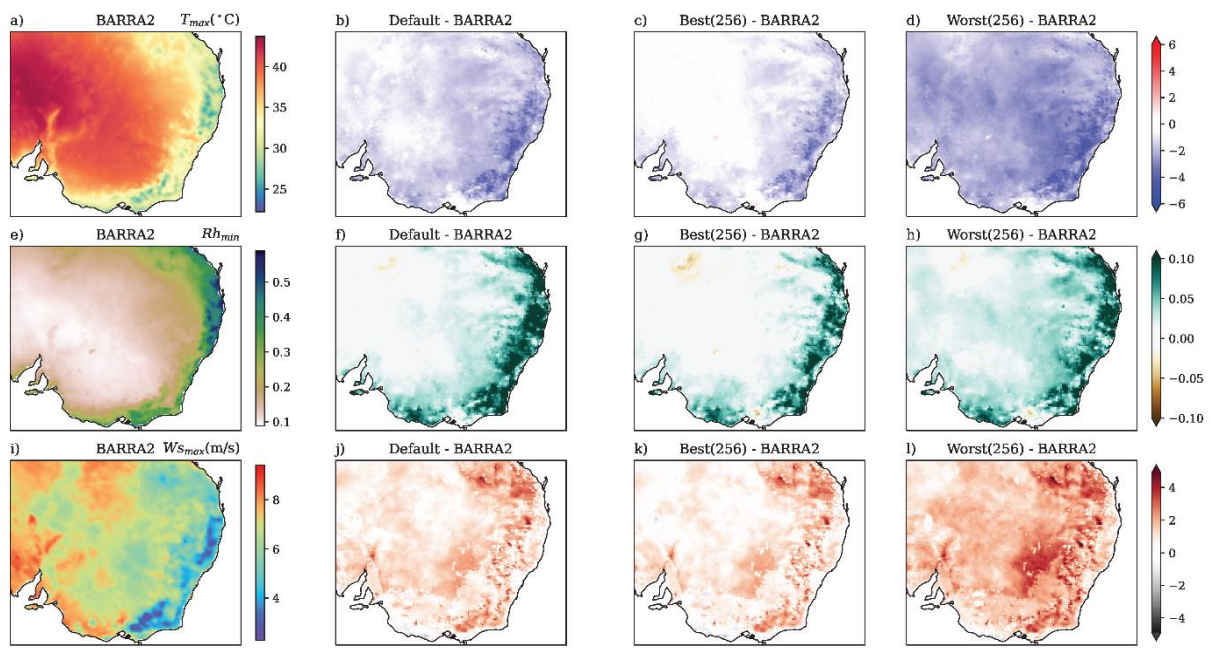

*Fig. 4 Spatial plot of average daily maximum temperature ($T_{max}$; °C) (a), daily minimum relative humidity ($Rh_{min}$) (e), and daily maximum wind speed ($Ws_{max}$; m/s) (i) during all days of both selected*

*events (2009 and 2019) using the BARRA2 data. Comparison of the WRF default parameters run (default), best (respective minimum RMSE of corresponding hourly meteorological variable data out of 256 runs) parameters run (best (256)), and worst (respective maximum RMSE of corresponding hourly meteorological variable data out of 256 runs) parameters run with respect to BARRA2 data for the considered meteorological variables. The mean bias of $T_{max}$ (b-d), $Rh_{min}$ (f-h), and $Ws_{max}$ (j-l) between default, best (256), and worst (256) runs with respect to BARRA2.*

### *3.4 Limitations*

We would like to emphasize that this study has a few caveats and limitations that should be noted. The parameter SA results are mostly dependent on regional conditions and could be on the type of the selected event (here heat extremes) (Chinta et al., 2021; Di et al., 2015; Ji et al., 2018; Quan et al., 2016). Hence, the SA results of this study may not be precisely applicable to other regions or other kinds of extreme events (for example, wet extremes). Testing the validity of this study's SA results for different extreme event types is beyond the scope of the present study.

Another limitation is that we have conducted the SA for only 24 parameters which may not be the complete list of all tuneable parameters in the WRF model. Nevertheless, this study is the first to systematically assess the sensitivity of WRF model parameters to considered meteorological variables during extreme heat events over southeast Australia. This study's results are mostly relevant for short simulations of these extremes, not necessarily for long-term regional climate model simulations. Nonetheless, our results will provide significant guidance for future research on parameter calibration of sensitive parameters for improving the WRF model simulation of heat extremes over southeast Australia. Future studies can use the present study's methodology or a similar one to conduct parameter SA of different regional atmosphere models in various regions for a variety of extreme events.

### *4 Conclusions*

Studies using regional atmospheric modelling systems to better understand extreme heat events such as heatwaves and bushfires rarely examine the model's sensitivity to tuneable parameters within parameterization schemes. This is the first study to address this issue over southeast Australia, a region that is highly susceptible to such extremes. We have investigated the sensitivity of 24 WRF tuneable parameters from seven different physics parameterisation schemes to the meteorological variables critical to the heat extremes over southeast Australia. The ML surrogate-based Sobol SA method is used to conduct the parameter SA during the two extreme heat events (2009 and 2019) in southeast Australia. Results show that only three parameters, P14 (scattering tuning parameter), P17 (multiplier of saturated soil water content), and P22 (profile shape exponent in the momentum diffusivity coefficient), are most important for the considered meteorological variables (T, Rh, and Ws) during heat extremes in southeast Australia. These results are consistent for both the considered events. These parameters are in the shortwave radiation, land surface, and planetary boundary layer schemes.

Further, we examined the physical significance of sensitive parameters on how they affect the simulation of considered meteorological variables. In conclusion, results showed that both low values of P14 and P17 simulate high T and low Rh, whereas low P14 and low P22 produce strong winds. Overall, this study's results suggest that only three out of 24 parameters needed to be considered for improving the simulation of heat extremes over southeast Australia. Future studies can focus on optimising these three sensitive parameters for improving the simulation of heat extremes over southeast Australia using multi-objective optimisation techniques. Our overall methodology will also be helpful for studies focussing on extreme heat events elsewhere.

## Acknowledgments

The authors would like to thank Chun-Hsu Su for providing the BARRA2 data. We acknowledge James Risbey's suggestion for improving the Figure 4. We would like to thank National Computing Infrastructure (NCI) Australia for providing computational resources. The authors would like to acknowledge the funding support of the CSIRO and Australian Climate Service.

*Supplementary material*

*Contents: - Figures S1-S4*

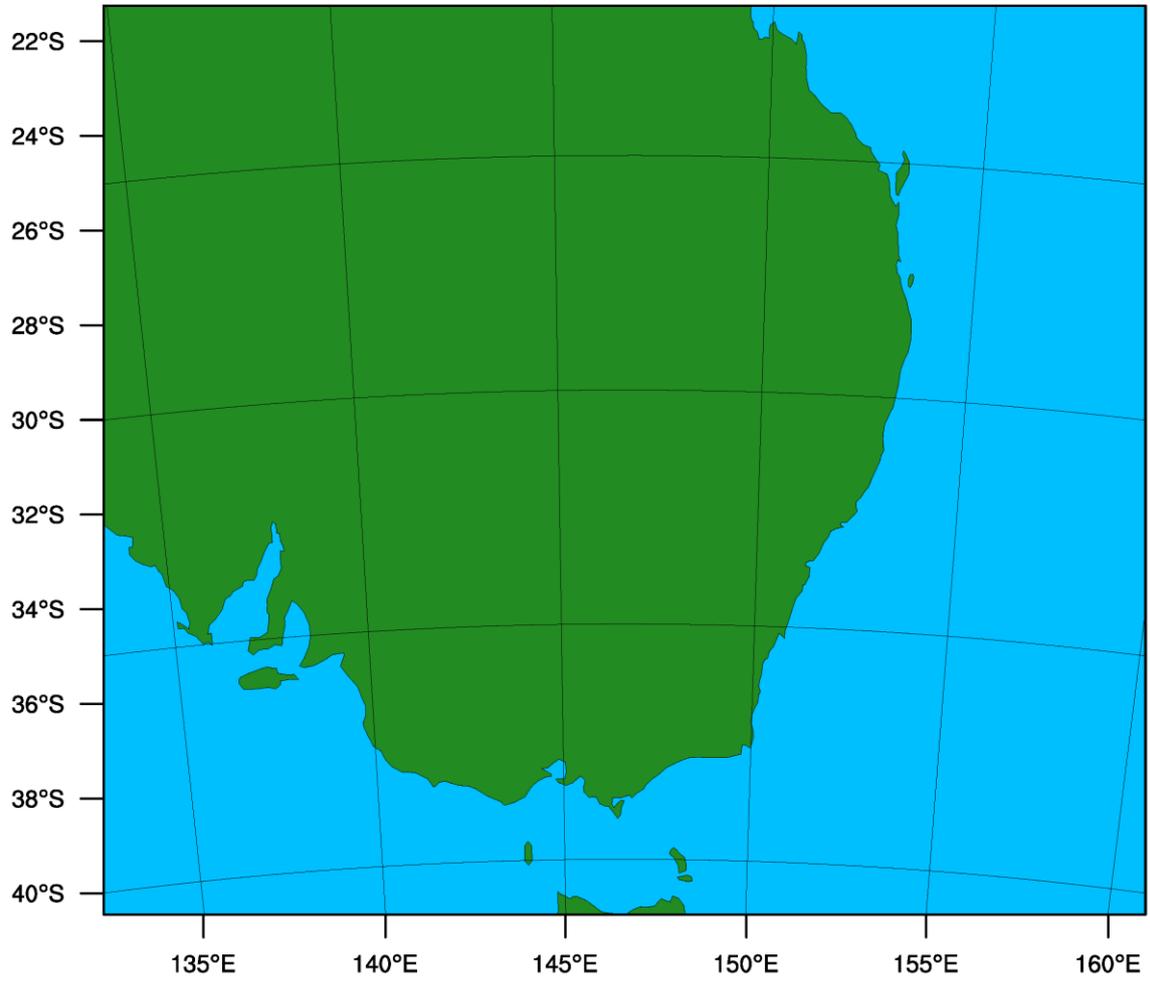

*Fig. S1 The WRF model domain with 12 km horizontal resolution over southeast Australia.*

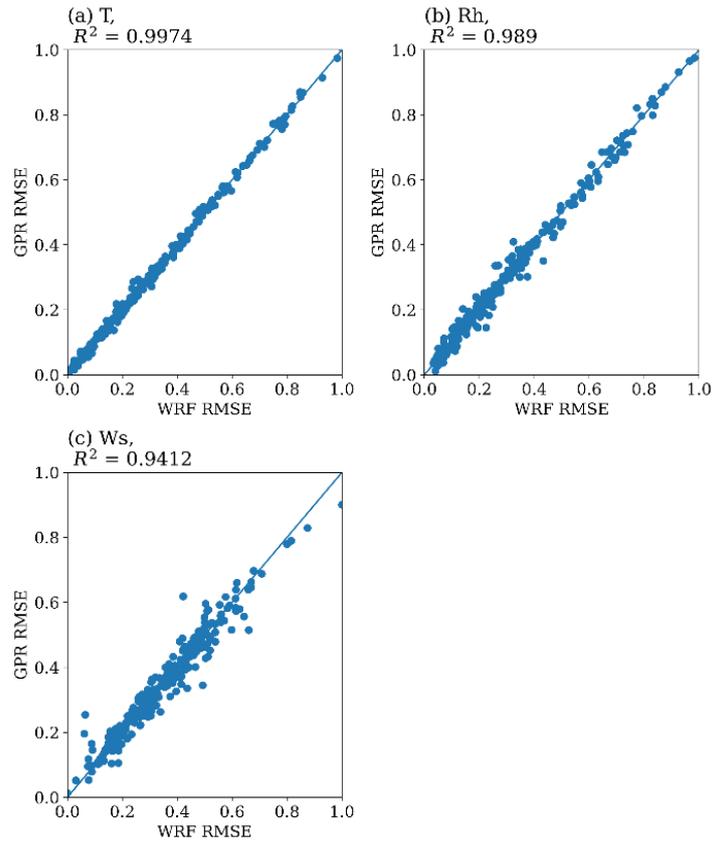

*Fig. S2 Comparison of normalised RMSE of air temperature (at 2m height) (T; °C), relative humidity (at 2m height) (Rh; %), and wind speed (at 10m height) (Ws; m/s) of 256 WRF parameter runs with the GPR predictions for the combined data of both considered events (2009 and 2019).*

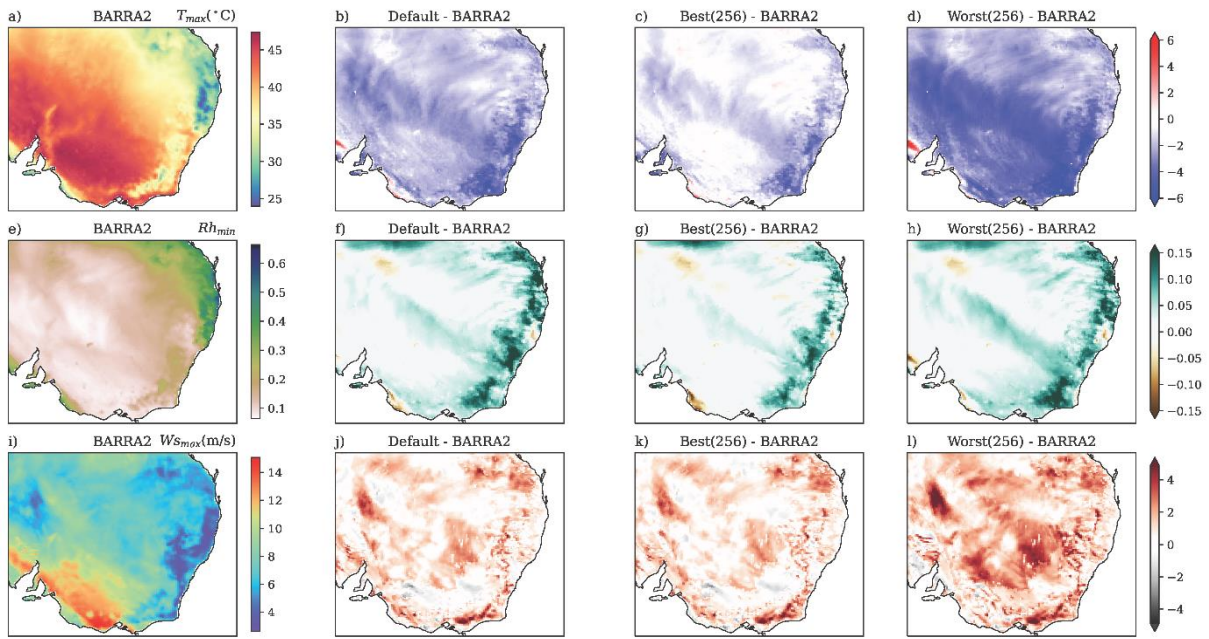

*Fig. S3 Spatial plot of daily maximum temperature ($T_{max}$; °C) (a), daily minimum relative humidity ($Rh_{min}$) (e), and daily maximum wind speed ($Ws_{max}$; m/s) (i) during the extreme hot day of 2009 event (07$^{th}$ Feb 2009) using the BARRA2 data. Comparison of the WRF default parameters run (default), best*

*(respective minimum RMSE of corresponding hourly meteorological variable data out of 256 runs) parameters run (best (256)), and worst (respective maximum RMSE of corresponding hourly meteorological variable data out of 256 runs) parameters run with respect to BARRA2 data for the considered meteorological variables. The mean bias of $T_{max}$ (b-d), $Rh_{min}$ (f-h), and $Ws_{max}$ (j-l) between default, best (256), and worst (256) runs with respect to BARRA2.*

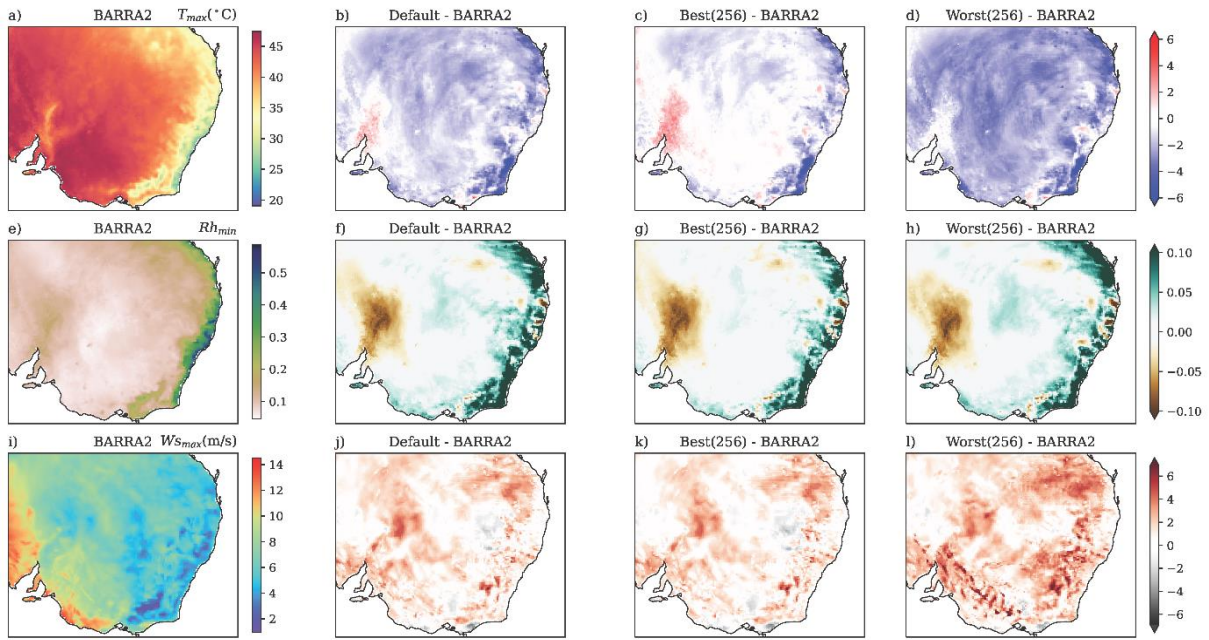

*Fig. S4 Same as figure S3 but for extreme hot day of 2019 event (20$^{th}$ Dec 2019).*